\documentclass{aa}
\usepackage{graphicx}

\begin{document}
\title{On the binding energy parameter of common envelope evolution}
\subtitle{Dependency on the definition of the stellar core boundary during spiral-in}

\author{T. M. Tauris\inst{1}
        \and
        J. D. M. Dewi\inst{2,3,4}
        }
\offprints{T. M. Tauris}
\institute{Nordic Institute for Theoretical Physics (NORDITA),
           Blegdamsvej 17, 2100 Copenhagen {\O}, Denmark\\
           \email{tauris@nordita.dk}
         \and Astronomical Institute {\it Anton Pannekoek},
              University of Amsterdam, Kruislaan 403, NL-1098 SJ Amsterdam,
              The Netherlands
         \and Bosscha Observatory, Lembang 40391, Bandung Indonesia
         \and Department of Astronomy, Institut Teknologi Bandung, 
              Jl. Ganesha 10, Bandung 40132, Indonesia\\
              \email{jasinta@astro.uva.nl}
}

\date{Received 9 November 2000 / Accepted 9 January 2001}

\abstract{
According to the standard picture for binary interactions, the outcome of 
binaries surviving the
evolution through a common envelope (CE) and spiral-in phase is determined
by the internal structure of the donor star at the onset of the mass transfer,
as well as the poorly-known efficiency parameter, $\eta _{\rm CE}$, for the ejection
of the {H}-envelope of the donor. In this Research Note we discuss the bifurcation point
which separates the ejected, unprocessed {H}-rich material from the inner core
region of the donor (the central part of the star which will later contract to
form a compact object). We demonstrate that the exact location of this point
is very important for evaluating the binding energy
parameter, $\lambda$, which is used to determine the post-CE 
orbital separation.\\
Here we compare various methods to define the bifurcation point
(core/envelope boundary) of evolved stars with masses 4, 7, 10 and 20$\,\mathrm{M}_{\odot}$.
We consider
the specific nuclear energy production rate profile, the change in the mass-density
gradient (Bisscheroux 1998), the inner region containing less than 10\% hydrogen, the method
suggested by Han et al.~(1994) and the entropy profile. 
We also calculated effective polytropic index profiles.\\
The entropy profile method measures the convective boundary (at the onset of
flatness in the specific entropy) which is not equivalent
to the core boundary for RGB stars. Hence, this method is not applicable for RGB stars,
unless the actual bifurcation point of a CE is located at
the bottom of the outer convection zone (resulting in larger values of 
$\lambda$ and larger post-CE orbital separations). 
On the AGB, where highly degenerate and condensed cores are formed,
we find good agreement between the 
various methods, except for massive ($\sim 20\,\mathrm{M}_{\odot}$) stars.
}

\maketitle


\section{Introduction}
The simple formalism introduced by Webbink (1984) and de~Kool (1990) to estimate
the orbital decay of binaries evolving through a CE and spiral-in evolution
requires knowledge about the core mass, $M_{\rm core}$, inside the bifurcation
point of the donor star as well as the $\lambda$-parameter, which is a numerical
factor introduced in order to correct their simple formula for estimating
the binding energy of the stellar envelope.
It has been common practice to use a constant value of $\lambda =0.5$
for the $\lambda$-parameter. We have demonstrated (Dewi \& Tauris 2000) that this 
is not a good approach, since $\lambda$ varies strongly throughout the evolution of
a star. 
We concluded that all observations of binary pulsars originating
from a CE evolution are consistent with $\eta_{\rm CE}\le1$, as a result
of large values of $\lambda$ possible when the internal thermodynamic energy is included.

We also raised the important question of {\em how} to define the core mass boundary
-- i.e. once the spiral-in process is initiated, and frictional torques deposit
kinetic energy in the envelope, where is the exact location of the point of bifurcation 
in the envelope which separates the ejected material from the remaining condensed
core region$\,$? This question is the key issue in this Research Note.
It is necessary to know $M_{\rm core}$
accurately, since it is used to determine the exact value of $\lambda$, which 
is a strongly increasing function exactly at the core/envelope transition.
Hence, the estimates of
the post-CE orbital separation also depend strongly on the value of $M_{\rm core}$.

We refer the reader to Dewi \& Tauris~(2000) and references therein for further 
details on the topic and the relevant energy equations and parameters.

\section{Structure profiles of evolved stars}
\begin{figure*}
        \begin{center}
         \centerline{\resizebox{17.4cm}{!}{\includegraphics{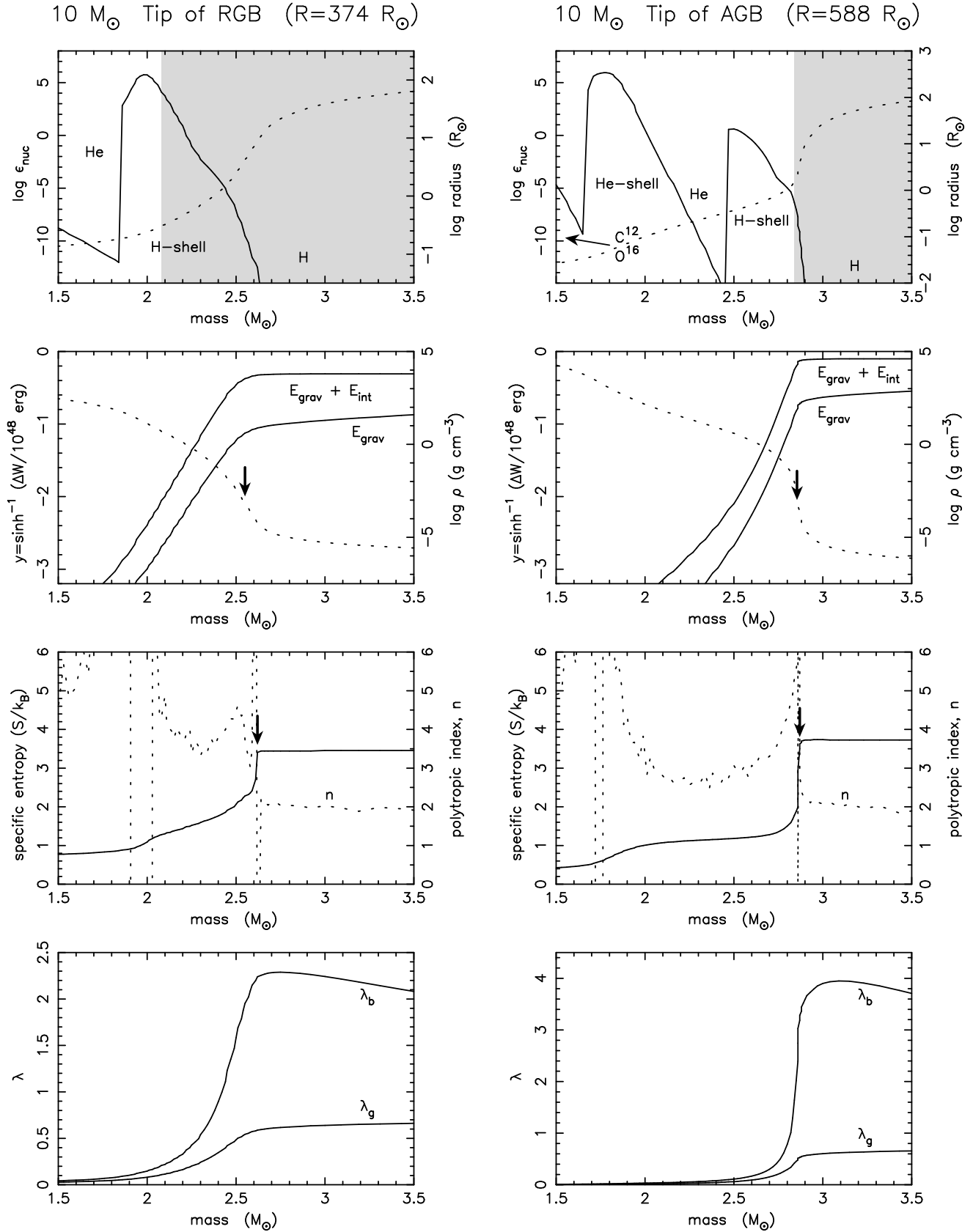}}}
         \caption[]{The internal structure of a 10$\,\mathrm{M}_{\odot}$ star 
                    at the tip of the RGB (left) and at the tip of the 
                    AGB (right). See text.}
         \end{center}
\end{figure*}

\setlength{\tabcolsep}{2.0pt}
\begin{table*}
         \caption[]{The values of $M_{\rm core}/\mathrm{M}_{\odot}$, $\lambda_{\mathrm g}$ and
          $\lambda_{\mathrm b}$ estimated from calculations of four different stars -- see text
          for discussion.}
         \begin{center}
         \begin{tabular}{lclllclll|clll|clllclll|clll}
          \cline{1-2}\cline{3-9}\cline{11-13}\cline{15-21}\cline{23-25}
          \noalign{\smallskip}
          & & \multicolumn{7}{c} {$M = 4.0\,\mathrm{M}_{\odot}$} & &
           \multicolumn{3}{c} {$M = 7.0\,\mathrm{M}_{\odot}$} & &
           \multicolumn{7}{c} {$M = 10.0\,\mathrm{M}_{\odot}$} & &
           \multicolumn{3}{c} {$M = 20.0\,\mathrm{M}_{\odot}$}\\
          \noalign{\smallskip}
          \cline{3-9}\cline{11-13}\cline{15-21}\cline{23-25}
          \noalign{\smallskip}
          & & \multicolumn{3}{c} {tip of RGB} & &
          \multicolumn{3}{c} {tip of AGB} & &
          \multicolumn{3}{c} {AGB} & &
          \multicolumn{3}{c} {tip of RGB} & &
          \multicolumn{3}{c} {tip of AGB} & &
          \multicolumn{3}{c} {tip of AGB}\\
          & & \multicolumn{3}{c} {R=67$\,\mathrm{R}_{\odot}$} & &
          \multicolumn{3}{c} {R=1040$\,\mathrm{R}_{\odot}$} & &
          \multicolumn{3}{c} {R=374$\,\mathrm{R}_{\odot}$} & &
          \multicolumn{3}{c} {R=374$\,\mathrm{R}_{\odot}$} & &
          \multicolumn{3}{c} {R=588$\,\mathrm{R}_{\odot}$} & &
          \multicolumn{3}{c} {R=1040$\,\mathrm{R}_{\odot}$}\\
          \noalign{\smallskip}
          \cline{3-5}\cline{7-9}\cline{11-13}\cline{15-17}\cline{19-21}\cline{23-25}
          \noalign{\smallskip}
          method & & $M_{\rm core}$ & $\lambda_\mathrm{g}$ & $\lambda_\mathrm{b}$ & &
          $M_{\rm core}$ & $\lambda_\mathrm{g}$ & $\lambda_\mathrm{b}$ & &
          $M_{\rm core}$ & $\lambda_\mathrm{g}$ & $\lambda_\mathrm{b}$ & &
          $M_{\rm core}$ & $\lambda_\mathrm{g}$ & $\lambda_\mathrm{b}$ & &
          $M_{\rm core}$ & $\lambda_\mathrm{g}$ & $\lambda_\mathrm{b}$ & &
          $M_{\rm core}$ & $\lambda_\mathrm{g}$ & $\lambda_\mathrm{b}$\\
          \noalign{\smallskip}
          \cline{1-2}\cline{3-5}\cline{7-9}\cline{11-13}\cline{15-17}\cline{19-21}\cline{23-25}
          \noalign{\smallskip}
          max $\epsilon _{\rm nuc}$ & & 0.58 & 0.32 & 0.62 & & 1.37 & 0.91 & --- & &
          1.58 & 0.10 & 0.20 & & 1.99 & 0.09 & 0.13 & & 2.47 & 0.06 & 0.10 & & 6.80 & 0.02 & 0.05\\
          $\mathrm{X}<0.10$ & & 0.59 & 0.36 & 0.70 & & 1.37 & 0.91 & --- & &
          1.80 & 0.59 & 2.85 & & 2.08 & 0.11 & 0.21 & & 2.84 & 0.40 & 1.80 & & 6.80 & 0.02 & 0.05\\ 
          $\partial ^2 \log \rho / \partial m^2 =0$& & 0.68 & 0.66 & 1.60 & & 1.37 & 0.91 & --- & &
          1.80 & 0.59 & 2.85 & & 2.54 & 0.54 & 1.80 & & 2.85 & 0.46 & 2.20 & & 7.50 & 0.22 & 0.95\\
          Han et al. & & 0.64 & 0.59 & 1.17 & & 1.37 & 0.91 & --- & &
          1.80 & 0.59 & 2.85 & & 2.52 & 0.51 & 1.75 & & 2.85 & 0.46 & 2.20 & & 7.60 & 0.32 & 1.70\\
          entropy profile& & 0.74 & 0.73 & 1.81 & & 1.37 & 0.91 & --- & &
          1.80 & 0.59 & 2.85 & & 2.62 & 0.60 & 2.23 & & 2.86 & 0.55 & 3.20 & & 7.80 & 0.45 & 3.50\\
          \cline{1-2}\cline{3-9}\cline{11-13}\cline{15-21}\cline{23-25}
         \end{tabular}
         \end{center}
\end{table*}
\setlength{\tabcolsep}{6pt}

In Fig.~1 we show an example of our calculations of the stellar structure for a 
10$\,\mathrm{M}_{\odot}$ star. The results for a 4, 7 and 20$\,\mathrm{M}_{\odot}$ star
are summarized in Table~1.
We used a chemical composition
of $\mathrm{X}=0.70$ and $\mathrm{Z}=0.02$, and a mixing-length parameter of
$\alpha =l/H_{\rm p}=2.0$. Convective overshooting was also taken into account.
Here we used an overshooting constant of $\delta_\mathrm{ov} = 0.10$ (Pols~et~al. 1998).
We obtained almost the same core masses using $\delta_\mathrm{ov} = 0.12$.
For the 10 and 20$\,\mathrm{M}_{\odot}$ star
we assumed a wind mass-loss rate according to de~Jager
(de~Jager~et~al. 1988; Nieuwenhuijzen \& de~Jager 1990). Hence, at 
the tip of the AGB these two stars are estimated to have evaporated down to
9.59 and 16.2$\,\mathrm{M}_{\odot}$, respectively. We assumed no mass loss from the
4 and 7$\,\mathrm{M}_{\odot}$ stars. 

We will now briefly discuss the calculations represented in Fig.~1.
In the bottom panels the two solid lines show our calculations of $\lambda$.
We distinguish $\lambda_{\rm b}$ (calculated from
the total binding energy) from  $\lambda_{\rm g}$ 
(as derived from gravitational binding energy alone) -- see Dewi \& Tauris~(2000).

\subsection{Energy production rate and chemical composition}
A very crude definition of the stellar core region is simply as the inner part 
containing less than a certain mass fraction of hydrogen. As an example, we used
$\mathrm{X}<0.10$ in Dewi \& Tauris~(2000) to define the core of the donor star. 
This is shown in the top panels of Fig.~1 as the area to the left of the gray shaded region.
The dotted line is the core radius as a function of its mass.\\ 
A somewhat more physically meaningful approach (although not necessarily more correct
in practice) is to define the core region below the (outer) shell burning source
-- for example, below the point of maximum energy production inside the H-burning shell.
In the top panels the solid line shows the local nuclear energy production 
rate. The shell burning regions are clearly seen. The most abundant 
local chemical elements are written in the panel. 
As can be seen from the figure and Table~1, this definition sometimes yields significantly
lower values of $M_{\rm core}$.

\subsection{The binding energy profile}
The solid lines in the second panel are $\sinh ^{-1} (\Delta W)$,
where $\Delta W = E_{\rm grav}+E_{\rm int}$ 
or $\Delta W = E_{\rm grav}$ 
is the binding energy of the envelope to the core (with or without the inclusion of
the internal thermodynamic energy, respectively). 
Han~et~al. (1994) introduced this function to define 
the core mass boundary.
The location is assumed to be at the transition between the strongly increasing
$\Delta W$ and the outer region where $\Delta W$ varies slowly with mass.
We used the intersection of straight-line fits to find $M_{\rm core}$. 

\subsection{The mass-density gradient}
The dotted line in the second panel shows the mass-density profile. The arrow indicates
the point where $\partial ^2 \log \rho / \partial m^2 = 0$.
This criterion was used by Bisscheroux~(1998) to locate the core boundary.
The disadvantage of this method, from a technical point of view, is
that there is not always a unique solution to
the equation (for small mass steps).

\subsection{The entropy profile and effective polytropic index}
The entropy profile for the $10\,\mathrm{M}_{\odot}$ star is plotted as a solid line in the third 
panel. Here ``entropy" refers to the local specific entropy per baryon in units of
the Boltzmann constant. The arrow indicates the sharp onset of the flat entropy gradient
-- another criterion for determining the bifurcation point.
The core masses at the tip of the RGB, defined by this method, are always larger than those 
derived from other criteria.
The clear discrepancy between the entropy profile method and the other 
alternatives on the RGB, leading to different estimates of the  
remaining mass after spiral-in, is easy to 
understand. The bifurcation point obtained from the entropy method is expected to be located
further out, since it is based on the transition between the convective and
the radiative layer in the stellar envelope. Unlike evolved low-mass ($1\,\mathrm{M}_{\odot}$) stars, 
these more massive stars do not have an outer convection zone which 
penetrates all the way down to the {H}-shell near the standard core boundary region
(e.g. Kippenhahn \& Weigert 1990). Hence, the entropy method results in 
(too) large a core mass on the RGB.

The dotted line in the third panel is the ``effective polytropic index" defined by:
$n\equiv 1/(\gamma -1)$ with the ``adiabatic" index: 
$\gamma =  \partial \ln P / \partial \ln \rho $.
Giant stars can be considered as a condensed polytrope with a core point mass
surrounded by an extended isentropic $n=3/2, \gamma=5/3$ envelope.
Less evolved stars are more realistically described as composite polytropes
consisting of $n=3, \gamma=4/3$ cores with $n=3/2, \gamma=5/3$ envelopes
(Hjellming \& Webbink 1987). Here, we are mostly interested in the clear
discontinuous behaviour of $n$ at the core boundary in order to determine 
its exact location. This discontinuity represents an important transition
in the stellar structure and could be interpreted as the
critical point of bifurcation at the end of the spiral-in and
envelope ejection phase (cf. Section~3.2).

\section{Discussion}
\subsection{Mass transfer initiated on the RGB vs. AGB}
We have now demonstrated how different methods result in different core masses
on the RGB and hence different values of $\lambda$. As an example, one can
consider a 4$\,\mathrm{M}_{\odot}$ star at the tip of the RGB. Here the maximum
energy production rate and the 10\% hydrogen criterion yield core masses
of 0.58 and 0.59$\,\mathrm{M}_{\odot}$, respectively. The mass-density gradient
method results in $M_{\rm core}=0.68\,\mathrm{M}_{\odot}$ and the outcome of
applying Han~et~al.'s method is approximately $0.64\,\mathrm{M}_{\odot}$. Using the entropy
profile and the effective polytropic index method yields 
$M_{\rm core}=0.74\,\mathrm{M}_{\odot}$. The spread in $M_{\rm core}$ from the
different approaches results in: $0.6<\lambda_{\rm b}<1.8$ and
$0.3<\lambda_{\rm g}<0.75$ (with and without internal thermodynamic energy).
If we now assume the simple CE energy equations (Webbink~1984; de~Kool~1990) 
to be approximately valid, and this $4\,\mathrm{M}_{\odot}$ ($R=67\,\mathrm{R}_{\odot}$) star is in
a common envelope with an in-spiralling $1.3\,\mathrm{M}_{\odot}$ neutron star, 
the evolution would then result in a final post-CE
orbital separation of 0.70 or 2.75$\,\mathrm{R}_{\odot}$,
for $M_{\rm core}=0.58$ and 0.74$\,\mathrm{M}_{\odot}$, respectively. Here we assumed 
$\lambda = \lambda_{\rm b}$, $\eta_{\rm CE}=1.0$ and no significant accretion
onto the neutron star during spiral-in. However, the core radius of the 
stripped star is 0.1$\,\mathrm{R}_{\odot}$ ($M_{\rm core}=0.58\,\mathrm{M}_{\odot}$) or 
3.5$\,\mathrm{R}_{\odot}$ ($M_{\rm core}=0.74\,\mathrm{M}_{\odot}$). Hence, in the latter
case, the large core would then have a radius larger than the predicted post-CE
separation. Therefore this system would coalesce during the spiral-in --
perhaps leaving behind a black hole.
Assuming $\lambda = \lambda_{\rm g}$ (i.e. considering gravitational binding
energy alone), or choosing a small efficiency parameter (e.g. $\eta _{\rm CE}=0.3$),
would decrease the predicted post-CE separation still further
and lead to an earlier coalescense. 
We conclude that knowledge of the accurate bifurcation boundary is crucial 
to forecast the outcome of a CE evolution on the RGB.

Many binaries come into contact while the donor star is expanding enormously on the AGB.
If we consider the same 4$\,\mathrm{M}_{\odot}$ star as before, but now
at the tip of the AGB, {\em all} methods yield {\em exactly} the same core mass.
For the 7 and 10$\,\mathrm{M}_{\odot}$ stars on the AGB,
the different methods also yield the same core mass (except for the location of 
the maximum energy production rate in the H-shell, which is below this boundary
at the outer edge of the shell source).\\
It is important to notice
the fact that the envelope binding energy is a strongly increasing
quantity near/at the core boundary on the AGB. Hence, the resulting values of $\lambda$ 
are also increasing sharply at this transition (see Fig.~1). Since the material 
of the donor star is peeled off from the outside, we suggest using
the (larger) value of lambda at the mesh point, just outside the bifurcation
point.

\subsection{Envelope ejection process}
While the location of the bifurcation point, separating the ejected envelope
from the condensed core region, is straightforward to define on the AGB, it is
important to know for RGB stars if the envelope of the evolved donor is
ejected at the bottom of the convection zone, or closer to the stellar center
at the core boundary
near/at the outer shell source (surrounding the inner part of the star).
For convective material in an isentropic
envelope, $R\propto M^{-1/3}$ and thus removal of envelope mass causes
the star to expand. Therefore, in a CE system, the envelope is easily peeled
off until the entropy profile increases outwards (in the deeper radiative layers). 
From this point on the star will no longer expand and the envelope may
already be separated from the inner region, depending
on the details of the ejection process.
It would be interesting to see if multi-dimensional
hydrodynamical calculations could answer this question.\\
Binaries which come into Roche-lobe contact early in the
stellar evolution of the donor will not survive the spiral-in process at all 
(there is simply not enough orbital energy available to expel the envelope).

The values of $\lambda$ presented in Table~1 in Dewi \& Tauris (2000)
were calculated from
the 10\% hydrogen criterion. From the work presented here we can see that
these values should therefore be taken as {\em minimum} values of $\lambda$,
supporting our previous conclusion that $\eta _{\rm CE}$ need not be large
since the value of $\lambda$ itself is expected to be large.

\begin{acknowledgements}
We thank the referee for improving (but substantially shortening) this paper.
T.~M.~T. acknowledges the receipt of a NORDITA fellowship.
This work was sponsored by NWO Spinoza Grant 08-0 to E.~P.~J. van~den~Heuvel. 
\end{acknowledgements}


\begin{thebibliography}{}
  \bibitem{} Bisscheroux B., 1998, M.Sc. Thesis, Univ. Amsterdam
  \bibitem{} de Jager C., Nieuwenhuijzen H., van der Hucht K. A., 1988,
             A\&AS 72, 259
  \bibitem{} de Kool M., 1990, ApJ 358, 189
  \bibitem{} Dewi J. D. M., Tauris T. M., 2000, A\&A 360, 1043
  \bibitem{} Han Z., Podsiadlowski P., Eggleton P. P., 1994, MNRAS 270, 121
  \bibitem{} Hjellming M. S., Webbink R. F., 1987, ApJ 318, 794
  \bibitem{} Kippenhahn R., Weigert A., 1990, Stellar Structure
             and Evolution, A\&A Library, Springer-Verlag
  \bibitem{} Nieuwenhuijzen H., de Jager C., 1990, A\&A 231, 134
  \bibitem{} Pols O. R., Schr\"{o}der K.-P., Hurley J. R., Tout C. A.,
             Eggleton P. P. , 1998, MNRAS 298, 525
  \bibitem{} Webbink R. F., 1984, ApJ 277, 355
\end{thebibliography}
\end{document}